\documentclass[10pt,aps,prd,twocolumn,showpacs,amsmath,amssymb,nofootinbib,eqsecnum,preprintnumbers,superscriptaddress]{revtex4-2}
\usepackage{amsmath,amssymb}
\usepackage{enumitem}  
\usepackage[usenames, dvipsnames]{color} 
\usepackage{graphicx} 
\usepackage{comment}
\usepackage[normalem]{ulem}
\usepackage[utf8]{inputenc}
\usepackage{url}
 \usepackage{xcolor}

\usepackage{hyperref}  

\usepackage{graphicx}
\usepackage{dcolumn}
\usepackage{bm}

\newcommand{\dd}{{\rm{d}}} 
\newcommand{\rovno}{\!\!& = &\!\!}
\newcommand{\be}{\begin{equation}}
\newcommand{\ee}{\end{equation}}
\newcommand{\ba}{\begin{eqnarray}}
\newcommand{\ea}{\end{eqnarray}}

\newcommand{\beq}{\begin{equation}}
\newcommand{\eeq}{\end{equation}}
\newcommand{\beqa}{\begin{eqnarray}}
\newcommand{\eeqa}{\end{eqnarray}}

\begin{document}

\title{Thermodynamics of Genuine Kerr--Bertotti--Robinson Black Holes}

\author{David Kubiz\v n\'ak}

\email{david.kubiznak@matfyz.cuni.cz}

\affiliation{Institute of Theoretical Physics, Faculty of Mathematics and Physics,
Charles University, Prague, V Hole{\v s}ovi{\v c}k{\' a}ch 2, 180 00 Prague 8, Czech Republic}

\author{Hryhorii Ovcharenko}

\email{hryhorii.ovcharenko@matfyz.cuni.cz}

\affiliation{Institute of Theoretical Physics, Faculty of Mathematics and Physics,
Charles University, Prague, V Hole{\v s}ovi{\v c}k{\' a}ch 2, 180 00 Prague 8, Czech Republic}

\author{Ji{\v r}{\'i} Podolsk{\'y}}

\email{jiri.podolsky@matfyz.cuni.cz}

\affiliation{Institute of Theoretical Physics, Faculty of Mathematics and Physics,
Charles University, Prague, V Hole{\v s}ovi{\v c}k{\' a}ch 2, 180 00 Prague 8, Czech Republic}

\date{August 23, 2026}

\begin{abstract}
We propose a thermodynamic description of the genuine
Kerr--Bertotti--Robinson (Kerr--BR) black hole spacetimes. Contrary to previous studies, such solutions no longer have electric charge. Consequently, the thermodynamics is governed by simple relations. Namely, upon introducing a suitable normalization factor for the timelike Killing vector, these spacetimes are characterized by the standard Hawking temperature, Bekenstein's entropy, Komar mass and angular momentum, and horizon angular velocity.
The corresponding mass is shown to obey the standard Christodoulou--Ruffini formula, along with the  standard 
first law and  Smarr relations. 
In fact, by a proper rescaling of coordinates and solution parameters, and up to a non-trivial parameter space constraint, we formally recover the thermodynamic quantities of the vacuum Kerr solution. This  is used to bring the genuine Kerr--BR spacetime into a Kerr-like form. 
\end{abstract}

 \maketitle

\section{Introduction}

Black hole thermodynamics is one of the most fascinating subjects in modern theoretical physics. It started
with an observation that black hole perturbations, governed by classical Einstein equations, obey the laws of black hole mechanics that are in many ways analogous to the laws of thermodynamics \cite{Bardeen:1973gs}. That this
is not a mere analogy but a true correspondence was confirmed in groundbreaking papers by Bekenstein
\cite{Bekenstein:1973ur} and Hawking \cite{Hawking:1975vcx}, who showed that when quantum effects are taken into account, black holes behave as
thermodynamic objects – they radiate away particles as a black body with temperature $T$, proportional to
the horizon surface gravity $\kappa$, and can be assigned an entropy $S$, identified with one quarter of the horizon
area, giving rise to a truly remarkable formula (in geometric units where $G_N=\hbar=k_B=c=1$):
\be \label{dictionary}
T=\frac{\kappa}{2\pi}\,,\qquad S=\frac{\mbox{Area}}{4}\,.
\ee
As of now, 
black hole thermodynamics 
is 
a well established scientific discipline that stands on several foundational pillars, such as {\em quantum field theory} in curved spacetime \cite{Birrell:1982ix}, {\em Euclidean action} calculations of the 
gravitational path integral \cite{Gibbons:1976ue}, 
or {\em covariant Noether's approach} \cite{Wald:1993nt}. 
It has triggered  many developments, including the advent of {\em gauge/gravity duality} \cite{Maldacena:1997re}, and 
provides key insights into the nature of {\em quantum gravity}, e.g. 
\cite{Ryu:2006bv}.

Despite the fact that general principles are well established, e.g., \cite{Hawking:1975vcx, Gibbons:1976ue, Wald:1993nt}, finding thermodynamic laws for concrete black hole spacetimes is a rather complicated 
task that requires a detailed analysis of such solutions. Foremost, to avoid degeneracy, one has to make sure that the resultant first law is of `{\em full cohomogeneity}', i.e., that the number of parameters characterizing the black hole solution matches the number of  black hole physical parameters that are varied in the first law. This may require expanding the  thermodynamic dictionary \eqref{dictionary} -- introducing 
{\em novel} thermodynamic
{\em quantities} that characterize such solutions.  
For example, apart from standard chemical and rotational charges, one may consider muduli scalar charges \cite{Gibbons:1996af};  in the AdS settings, it is reasonable to introduce the notion of a 
{\em thermodynamic volume} \cite{Kastor:2009wy, Dolan:2010ha,Cvetic:2010jb}; the presence of cosmic strings (as in the  C-metric) gives rise to {\em thermodynamic length} \cite{Appels:2017xoe, Anabalon:2018qfv}; 
for the Misner string spacetimes, one may introduce the corresponding {\em Misner charges}
\cite{Bordo:2019tyh}; and for black holes in an external Bonnor--Melvin-type magnetic field $B$, one may consider the corresponding 
\be \label{mudB}
-\mu\, \delta B
\ee 
term, 
where $\mu$ is the {\em magnetic moment} of the system
\cite{Gibbons:2013dna}. (See,  however, an alternative proposal for the latter case, where this term is not considered and the resultant first law is of higher than full cohomogeneity \cite{Astorino:2016hls}.)

Another crucial step in deriving the thermodynamic description is to determine the   {\em thermodynamic mass}. While in the AdS settings, there exists a well-established covariant  procedure for finding such $M$ (one can, for example, use the 
holographic \cite{Henningson:1998gx} or topological \cite{Aros:1999id} renormalization, or the conformal methods \cite{Ashtekar:1999jx}), 
there is no universally agreed method
for defining the mass of spacetimes with more complicated  asymptotics. In fact, it is not even obvious
that the mass of such a spacetime should be well-defined \cite{Booth:2015nwa}. Nevertheless, it is sometimes possible to obtain the thermodynamic  mass by imposing the corresponding first law and Smarr relations, or by using some additional assumptions. In what follows, we shall adopt  such an approach.

Recently, a fascinating novel class of magnetized black holes  has been discovered  -- the so called (original) {\em Kerr--Bertotti--Robinson} (Kerr--BR) spacetimes \cite{Podolsky:2025tle, Ovcharenko:2025cpm}. Such solutions 
are {\em astrophysically} very interesting as they can describe black holes immersed in magnetic fields of the surrounding plasma, while avoiding a number of undesired  features of the Bonnor--Melvin black hole spacetimes, e.g. \cite{Gibbons:2013yq}.
Although the properties of such solutions are  gradually uncovered, see e.g. \cite{Gray:2025lwy, DiPinto:2026ynu, Zhou:2026tkm},
their 
thermodynamic description, so far, remains 
to be properly 
understood; see, however, the recent proposals \cite{Astorino:2025lih, Hu:2026slp}. Namely, in  \cite{Astorino:2025lih} (and also \cite{DiPinto:2026ynu}) only a non-rotating  case of {\em Schwarschild--Bertotti--Robinson} (Schwarzschild--BR) black holes  was considered. At the same time, the reference \cite{Hu:2026slp} used 
the original (charged) Kerr--BR solution of \cite{Podolsky:2025tle, Ovcharenko:2025cpm}, where the external magnetic field parameter $B$ is intertwined 
with a non-trivial electric charge of the solution, resulting in {\em constrained} thermodynamics.
Most recently, the two contributions were disentangled in \cite{OPnew}, and the {\em genuine} (uncharged) Kerr--BR solution was identified.

It is the aim of this paper to study the thermodynamics  of the genuine Kerr-BR black hole. We start by reviewing the corresponding solution.

\section{Genuine Kerr--BR solution}
The genuine Kerr--BR  metric found in \cite{OPnew}, representing an uncharged rotating black hole immersed in an external magnetic field $B$, takes the following form:
\ba 
\dd s^2 \rovno \frac{1}{\Omega^2}\Bigl[-\frac{Q}{\rho^2}\Bigl(\frac{\dd t}{\beta}-a\sin^2\, \theta\frac{\dd\phi}{K}\Bigr)^2+\frac{\rho^2}{Q}\dd r^2+\frac{\rho^2}{P}\dd \theta^2\nonumber\\
&&\qquad +\frac{P}{\rho^2}\sin^2\!\theta\Bigl(\frac{a\, \dd t}{\beta}-(r^2+a^2)\frac{\dd\phi}{K}\Bigr)^2\,\Bigr]\,,
\label{KBR0metric}
\ea 
with the Maxwell field potential given by 
\ba 
A \rovno -\frac{B}{\Omega\, \rho^2}\Bigl((1+m^2B^2)\rho^2-mr(1+\cos^2\!\theta)\Bigr)\frac{a\,\dd t}{\beta}\nonumber\\
&&+\frac{B}{\Omega\, \rho^2}\Bigl(a^2(1+m^2B^2){\rho^2}-mr(a^2-r^2)\sin^2\!\theta\Bigr)\frac{\dd\phi}{K}\nonumber\\
&&+\frac{\Omega-1}{\Omega\, B}\frac{\dd\phi}{K}\,.\label{A_1_form}
\ea 

Requiring $\phi$ to be periodic with period $2\pi$, ${\phi\in[0, 2\pi)}$,  we have introduced a conicity parameter ${K>0}$ (in the notations of \cite{OPnew} ${K=1/C}$). Moreover, we have introduced a constant factor~$\beta$, accounting for a possibly different normalization of the timelike Killing vector~$\partial_t$. 
The remaining metric functions are given by 
\ba 
\rho^2 \rovno r^2+a^2\cos^2\!\theta\,, \nonumber\\
P \rovno 1+B^2\bigl[m^2-a^2(1+m^2B^2)^2\bigr]\cos^2\!\theta\,, \nonumber\\
Q \rovno I\,\Delta\,,\qquad \Omega^2 = I-B^2\Delta \cos^2\!\theta \,,\nonumber\\
I \rovno (1-mB^2r)^2+B^2r^2\,,\nonumber\\
\Delta \rovno (r^2+a^2)(1+m^2B^2)-2mr\,,\label{Delta}
\ea 
where ${P\ge0}$, ${I>0}$, and $m$ and $a$ are the mass and rotation parameters, respectively. 

As discussed in \cite{OPnew}, and contrary to the `old version' of Kerr--BR black hole \cite{Podolsky:2025tle, Ovcharenko:2025cpm}, the spacetime is now completely ``charge free'', as we have 
\be 
Q_e=\frac{1}{4\pi}\int_{S^2} *F=0\,,\qquad Q_m=\frac{1}{4\pi}
\int_{S^2} F=0\,,
\ee 
over the spheres at any~$r$. 
This is thus a genuine electro-vacuum solution with source free Maxwell equations.

Moreover, as derived in \cite{OPnew}, cf. Eq.~(4.11), the spacetime is free of conical singularities on both axes, provided
\ba 
K = b^2\,K_a \,,
\ea  
where the constants are defined as
\ba 
b   \rovno \sqrt{1+m^2B^2}\,,\nonumber\\
K_a \rovno 1-a^2b^2B^2\,.
\ea  
Interestingly, using such ${K_a>0}$, we get
\ba 
I \rovno K_a+B^2\Delta\,,\nonumber\\
\Delta \rovno (r^2+a^2)\,b^2-2mr\,,
\ea 
implying very nice and compact expressions for the metric functions in \eqref{KBR0metric},
\ba 
P \rovno \sin^2\!\theta + K \cos^2\!\theta \,,\nonumber\\ 
Q \rovno \Delta \,(K_a+B^2\Delta)\,,\nonumber\\
\Omega^2 \rovno K_a + B^2 \Delta \sin^2\!\theta\,.\label{Om_2_simp}
\ea 
We can also rewrite the vector potential \eqref{A_1_form} in the `frame form' as
\be
A = \frac{A_1}{\rho^2}\Bigl(\frac{\dd t}{\beta}-a\sin^2\, \theta\frac{\dd\phi}{K}\Bigr)+\frac{A_2}{\rho^2}\Bigl(\frac{a\, \dd t}{\beta}-(r^2+a^2)\frac{\dd\phi}{K}\Bigr)\,,
\ee 
where
\ba 
A_1 \rovno  \frac{B}{\Omega}\,a r\,(2m-b^2\,r)+a\,\frac{\Omega-1}{\Omega B}\,,\nonumber\\
A_2 \rovno -\frac{B}{\Omega}\,(a^2b^2\cos^2\!\theta+mr\sin^2\!\theta)-\frac{\Omega-1}{\Omega B}\,.
\ea 

 For ${B=0}$, implying ${b=1=K_a}$, the Maxwell field vanishes, and the metric reduces to the Kerr black hole in the usual Boyer--Lindquist coordinates.

\section{Thermodynamics}
Let us now proceed to the thermodynamic description and properties of these black holes.

As always, the simplest characteristics to determine are the ``horizon quantities'', such as the Hawking temperature~$T$, the Bekenstein entropy~$S$, and the horizon angular velocity~$\Omega_+$. 
The (outer) horizon is located at~$r_+$, obtained as the largest root ${\Delta(r_+)=0}$ of \eqref{Delta}.\footnote{In black hole thermodynamics, the horizon equation is 
used to eliminate one parameter.  Typically, the simplest approach is to eliminate the mass parameter $m$, as $\Delta$ is usually linear in $m$. However, in our instance, this is not the case, and in this section, we shall solve this equation either for $r_+$ or for $a^2$. In the next section, we shall re-parametrize the coordinates and parameters of the solution so that the horizon equation becomes linear in the mass parameter.}
\be \label{De}
2m\, r_+ = (1+m^2B^2)(r_+^2+a^2)\,.
\ee 
It is a Killing horizon, generated by the Killing field
\be 
\xi=\partial_t +\Omega_+\, \partial_\phi\,,\label{xi_def}
\ee 
where the horizon angular velocity is given by 
\be 
\Omega_+=\omega|_{r=r_+}\,,\qquad \omega=-\frac{g_{t\phi}}{g_{\phi\phi}}\,.
\ee 
This yields a simple formula 
\be \label{Omega}
\Omega_+=\frac{K\,a}{\beta\,(r_+^2+a^2)}\,,
\ee 
and one can easily check that ${\xi^2|_{r=r_+}=0}$.

The Hawking temperature is  determined by the horizon surface gravity $\kappa$ as ${T=\kappa/(2\pi)}$, and reads 
\ba \label{T}
T \rovno \frac{1}{4\pi \beta}\,\frac{Q'(r_+)}{r_+^2+a^2}=\frac{K_a}{4\pi \beta}\,\frac{\Delta'(r_+)}{r_+^2+a^2}\nonumber\\
\rovno \frac{K}{4\pi\beta\, r_+}\,\frac{r_+^2-a^2}{r_+^2+a^2}\,,
\ea 
upon using the horizon equation \eqref{De}.
It can now be immediately seen that the temperature of the \emph{extreme} genuine Kerr--BR black hole is \emph{zero} because its degenerate horizon is given by ${r_+=a}$, see \cite{OPnew}.

The horizon area is
\be 
{\mbox{Area}}= \!\int\!\! \sqrt{g_{\theta \theta}\, g_{\phi\phi}} \,\dd\theta \dd\phi=\frac{2\pi (r_+^2+a^2)}{K}\!\int \frac{\sin\theta\, \dd\theta}{\Omega^2(r_+)}\,.
\ee  
The last calculation is aided by the fact that on the horizon, $\Omega^2(r_+)$ is constant, given by ${\Omega^2(r_+)=K_a}$. Applying Bekenstein's area law then yields the following entropy:
\be \label{S}
S=\frac{{\mbox{Area}}}{4}
=\frac{\pi b^2}{K^2}\,(r_+^2+a^2)\,. 
\ee

Next, we wish to calculate the angular momentum $J$. To this purpose, we simply use the Komar integration\footnote{It is well known that in spacetimes that are not asymptotically flat, the Komar integral for the mass often yields incorrect or divergent expressions. However, for a calculation of the angular momentum, it should be sufficient, and we proceed to use it; see also Appendix~\ref{App} for the use of `upgraded' Komar integration.}
\be 
J=\frac{1}{16\pi}\int_{\Sigma_2} *d \eta\,,
\ee 
where $\eta$ is the axial Killing vector, $\eta=\partial_\phi$\,.
The problem here is to determine the co-dimension two hypersurface $\Sigma_2$, over which one should integrate. One can think that the (conformal) infinity is, similar to the C-metric, e.g., \cite{Anabalon:2018qfv},  determined by the vanishing conformal factor ${\Omega=0}$. However, for the genuine Kerr--BR spacetime this is actually {\em not} the case. For example, from (\ref{Om_2_simp}) one can see that for generic $\theta$, $\Omega^2$ is positive. Thus, the problem of finding the correct conformal boundary of genuine Kerr--BR spacetime is much more complicated, and we postpone it to future works (see, however,   \cite{Zhou:2026tkm} for a very recent discussion of the global structure of the ``old Kerr--BR'' solution). Here we simply assume that it happens for large~$r$, and calculate the above integral over the ``sphere at infinity''. This then straightforwardly yields
\be \label{J}
J=\frac{m\,a}{K^2}\,.
\ee 
In what follows, we identify the corresponding conjugate quantity with $\Omega_+$.\footnote{We ignore here the possible contribution from $\omega$ at infinity, as such a quantity is $\theta$-dependent. For example, at the poles, we find  
\be 
\omega_{\infty}|_{\theta=0,\pi}=\frac{aB^2K(1+B^2m^2)}{\alpha (m^2a^2B^4+a^2B^2-1)}\,,
\ee 
while a slightly more complicated expression is obtained in the equatorial plane. 
Whether this assumption is truly valid remains to be seen.  }

The last quantity to determine is the mass of the hole. Since we are not in a spacetime with standard asymptotics, 
this is not an easy task.
We shall determine it in two independent  (and supplementary) ways: i) by assuming the standard {\em Christodoulou--Ruffini}  (CR) formula, and ii) by using the {\em upgraded Komar integration} together with the {\em standard} first law.

i) Similar to the recent work \cite{Hu:2026slp}, we may assume the validity of the standard CR  formula:\footnote{Note that such a formula need not be valid in a general spacetime. For example, already in the AdS case, it picks up modifications, e.g. \cite{Caldarelli:1999xj}; see also \cite{Gregory:2019dtq} for the AdS C-metric  extension.}
\be\label{CR} 
M^2=\frac{S}{4\pi}+\pi\, \frac{J^2}{S}\,.
\ee 
It follows that the conjugate quantities 
are given by 
\ba\label{TO}  
T_{\mbox{\tiny CR}}\rovno\Bigl(\frac{\partial M}{\partial S}\Bigr)_{\!J}=\frac{1}{8\pi M}\Bigl(1-4\pi^2 \frac{J^2}{S^2}\Bigr)\,,\nonumber\\
\Omega_{\mbox{\tiny CR}}\rovno\Bigl(\frac{\partial M}{\partial J}\Bigr)_{\!S}=\frac{\pi J}{MS}\,.
\ea 
Evaluating the right hand side of \eqref{CR} using the derived formulas \eqref{S} and \eqref{J} for entropy and angular momentum, applying the relation \eqref{De},  yields the mass $M$ in the very simple form 
\be\label{M} 
M=\frac{m}{K b}\,.
\ee 
Moreover, using the formulas in \eqref{TO}, we find agreement of $T_{\mbox{\tiny CR}}$ and $\Omega_{\mbox{\tiny CR}}$ with the previously derived $T$ and $\Omega_+$ given by \eqref{T} and \eqref{Omega}, respectively, provided we identify $\beta$ as: 
\be \label{Mbeta}
\beta=b=\sqrt{1+m^2B^2}\,.
\ee 
This completely fixes all thermodynamic quantities.
Let us also notice that such $\beta$ is \emph{independent} of~$a$. In fact, the same $\beta$ was already considered in \cite{Astorino:2025lih} for constructing the thermodynamics of the Schwarzschild--BR black holes (see also \cite{DiPinto:2026ynu}).

One can now check by a direct calculation that the  {\em standard first law} and the {\em Smarr relation} are valid, namely
\ba 
\delta M \rovno T\,\delta S+\Omega_+ \delta J\,,\label{first}\\
M\rovno 2\,(TS+\Omega_+ J)\,.\label{Smarr}
\ea 
Here, actually, \emph{3 physical parameters can be varied}, e.g. $\{m, a, B\}$. Interestingly, this makes the first law of  {\em more than} ``full cohomogeneity''. In deriving  \eqref{first} we eliminated $r_+(m, a, B)$ by using the relation \eqref{De}, that also implies
\be 
\delta r_+ = \frac{K}{4\pi\beta\,T\, m \,r_+}\big[\,r_+\,\delta m -\beta^2a\,\delta a
 -(r_+^2+a^2)\, \beta \,\delta\beta \,\big]\,. 
\ee 
Note, however,  that sometimes it may be reasonable to treat the external field $B$ as fixed, and vary only the parameters $m$ and $a$.

ii) Alternatively, we may construct the thermodynamics as follows. We may {\em assume} $M$ to be proportional to the mass parameter $m$, and inversely proportional to $K$ and~$\beta$, similar to  \cite{Anabalon:2018qfv}. The simplest such ansatz is
\be\label{ansatz}
M=\frac{m}{K \beta}\,,
\ee
where the ``normalization factor'' $\beta$ remains to be determined. In fact, this ansatz for $M$  exactly coincides with the one obtained by the upgraded Komar integration  discussed in the Appendix~\ref{App}. 
Eliminating $r_+$ via \eqref{De}, $\beta$ is of the form  
\be\label{beta} 
\beta=\beta(m,a,B)\,.
\ee 
By imposing the first law \eqref{first}, we obtain 3 partial differential equations for $\beta$, which fix $\beta$ to the form \eqref{Mbeta}
up to a constant factor that can be set to unity  by considering the ${B\to 0}$ limit. Once the parameter $\beta$ is fixed by the first law, it is easy to verify that the above quantities then satisfy both the Smarr relation \eqref{Smarr} and the CR formula \eqref{CR}.

\section{New form of the Kerr--BR metric and its Kerr-like thermodynamics}

In the previous section, we were able to construct a consistent thermodynamic description of the genuine Kerr--BR black hole, obeying all relations (standard first law, the Smarr relation, and the CR formula) provided  we identify the scaling parameter $\beta$ with $b$ according to the equality  \eqref{Mbeta}.  As we shall now show, such thermodynamics is, in fact, {\em formally identical} to that of standard Kerr.

To see this, let us 
introduce a new set of parameters $\{M,\tilde{a},\tilde{r}_+\}$ instead of $\{m,{a}, {r}_+\}$, defined as  
\ba 
m=K b \,M\,,\qquad 
a=\frac{K}{b}\,\tilde{a}\,,\qquad 
r_+=\frac{K}{b}\,\tilde{r}_+\,.
\ea 
The horizon equation \eqref{De} then becomes {\em linear} in the physical mass $M$, namely we get  ${2M \tilde{r}_+=\tilde{r}_+^2+\tilde{a}^2}$, and we obtain the following thermodynamic quantities: 
\ba\label{KerrTDs} 
M \rovno \frac{\tilde{r}_+^2+\tilde{a}^2}{2\tilde{r}_+}\,,\qquad 
J=M\tilde{a}\,,\qquad 
\Omega_+=\frac{\tilde{a}}{\tilde{r}_+^2+\tilde{a}^2}\,,\nonumber\\
S \rovno \pi\, (\tilde{r}^2_++\tilde{a}^2)\,,\qquad 
T=\frac{1}{4\pi \tilde{r}_+}\,\frac{\tilde{r}^2-\tilde{a}^2}{
\tilde{r}_+^2+\tilde{a}^2}
\,,
\ea 
which are {\em formally} completely the same as those of Kerr in the standard Boyer--Lindquist coordinates.

The above simplification of the thermodynamic quantities naturally raises the question of whether it is also  possible to rewrite the genuine Kerr--BR metric in terms of some new coordinates and parameters so that it becomes ``closer to'' the Kerr spacetime. Indeed, this is the case. If we define
\ba 
r=\frac{K}{\beta}\,\tilde{r}\,,\qquad
\rho^2=\frac{K^2}{\beta^2}\,\tilde{\rho}^2\,,\qquad \Delta=K^2\,\tilde{\Delta}\,,\nonumber\\
P=K\,\tilde{P}\,,\qquad
Q=\frac{K^3}{\beta^2}\,\tilde{Q}\,,\qquad 
\Omega^2=\frac{K}{\beta^2}\,\tilde{\Omega}^2\,,
\ea 
and
\be 
\tilde{B}^2=K\beta^2 B^2\,,
\ee
the genuine Kerr--BR metric \eqref{KBR0metric} takes the following ``Kerr-like'' form:
\ba 
\dd s^2 \rovno \frac{1}{\tilde{\Omega}^2}\Bigl[-\frac{\tilde{Q}}{\tilde{\rho}^2}\Bigl(\dd t-\tilde{a}\sin^2\!\theta \dd\phi\Bigr)^2+\frac{\tilde{\rho}^2}{\tilde{Q}}\dd \tilde{r}^2+\frac{\tilde{\rho}^2}{\tilde{P}}\dd \theta^2\nonumber\\
&&\qquad +\frac{\tilde{P}}{\tilde{\rho}^2}\sin^2\!\theta\Bigl({\tilde{a}\, \dd t}-(\tilde{r}^2+\tilde{a}^2){\dd\phi}\Bigr)^2\,\Bigr]\,,
\ea 
where
\ba\label{newFunctions} 
\tilde{\rho}^2 \rovno \tilde{r}^2+\tilde{a}^2\cos^2\!\theta\,, \nonumber\\
\tilde{P} \rovno 1-\tilde{B}^2(M^2-\tilde{a}^2)\sin^2\!\theta\,, \nonumber\\
\tilde{Q} \rovno \tilde{\Delta}\,(1+\tilde{B}^2\tilde{\Delta})\,,\nonumber\\
\tilde{\Omega}^2 \rovno 1+\tilde{B}^2\tilde{\Delta} \sin^2\!\theta\,,\nonumber\\
\tilde{\Delta} \rovno \tilde{r}^2-2M\,\tilde{r}+\tilde{a}^2\,.
\ea 
Interestingly, on both axes ${\theta=0, \pi}$ there is ${\tilde{P}=1=\tilde{\Omega}}$. Obviously, no conical deficits thus occur, and we simply have the standard angular range ${\phi\in [0,2\pi)}$. 

Moreover, the horizons are given by the condition ${\tilde{\Delta}=0}$, which gives 
\be \label{De2}
 \tilde{r}_+ = M+\sqrt{M^2-\tilde{a}^2}\,.
\ee 
This is the same expression as for the Kerr black hole. Also, there is a constraint ${\tilde{a}\le M}$ to avoid a naked singularity.

Finally, the 4-potential in this parameterization takes the form
\begin{align}
A=\dfrac{\tilde{A}_1}{\tilde{\rho}^2}\big(\dd t-\tilde{a} \sin^2\theta\,\dd \phi\big)+\dfrac{\tilde{A}_2}{\tilde{\rho}^2}\big(\tilde{a}\dd t-(\tilde{r}^2+\tilde{a}^2)\,\dd \phi\big)\,,
\end{align}
where
\ba
    \tilde{A}_1 \rovno \dfrac{\tilde{a} \tilde{B}}{\tilde{\Omega}}\, \tilde{r}\, (2M-\tilde{r})\nonumber\\
    && 
    +\dfrac{\tilde{a}}{\tilde{\Omega} \tilde{B}}\Big[\sqrt{1+\tilde{a}^2\tilde{B}^2}\,\tilde{\Omega}-(1+\tilde{a}^2\tilde{B}^2)\Big]\,,\nonumber\\
    \tilde{A}_2 \rovno -\dfrac{\tilde{B}}{\tilde{\Omega}}\,(\tilde{a}^2\cos^2\theta+M \tilde{r} \sin^2\theta)\nonumber\\
    && -\dfrac{1}{\tilde{\Omega} \tilde{B}}\,\Big[\sqrt{1+\tilde{a}^2\tilde{B}^2}\,\tilde{\Omega}-(1+\tilde{a}^2\tilde{B}^2)\Big]\,.
\ea

\section{Behavior of free energy}

Although in this new form of the metric the thermodynamics is now formally identical to that of Kerr, there is a subtle difference. Namely, one has to take into account the parameter space {\em constraint}  coming from the positivity of  $\tilde{P}$, as given by \eqref{newFunctions}, namely that\footnote{similar types of ``external'' constraints affecting thermodynamics were, for example, considered in \cite{Abbasvandi:2018vsh, Abbasvandi:2019vfz}.}
\be \label{constr}
\tilde{B}^2(M^2-\tilde{a}^2)\leq 1\,.
\ee 
To see how this constraint affects the thermodynamics, we shall display  the corresponding free energy in the canonical and grandcanonical ensembles. Without loss of generality, we consider $\tilde B\geq 0$. 

\subsection{Canonical ensemble}
In the canonical (fixed $J$) ensemble, the corresponding free energy reads:
\be 
F\equiv F(T,J)=M-TS=\frac{\tilde{r}_+^2+3\,\tilde{a}^2}{4\,\tilde{r}_+}\,,
\ee 
with $T$ and $J$ given parametrically in terms of $\tilde{a}$ and $\tilde{r}_+$ via the relations \eqref{KerrTDs}.

\begin{figure}
    \centering
\includegraphics[width=0.45\textwidth]{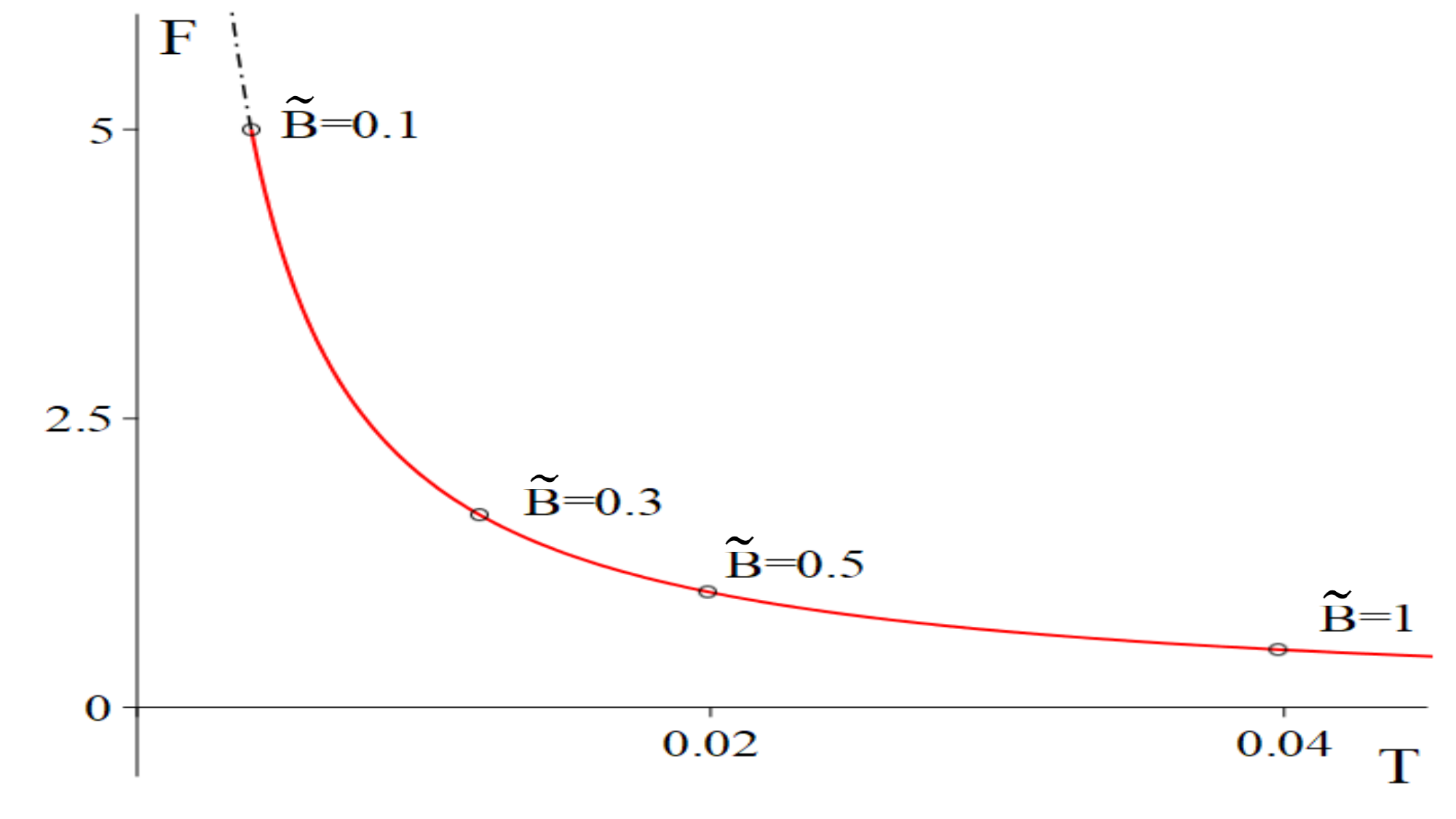}
\caption{{\bf Free energy of the Schwarzschild--BR black hole}, displayed as a function of temperature $T$ for several values of the magnetic field $\tilde B$. Although formally identical to that of Schwarzschild, there is now a {\em maximum for} ${1/(2\tilde B)}$ (and the corresponding minimum of $T$). These are displayed by the black circles for ${\tilde B=1, 0.5, 0.3, 0.1}$ (from right to left) where the red curve $F(T)$ (emanating from the right) terminates. In particular, for ${\tilde{B}=0.1}$ this is displayed by the uppermost black circle; the dashed black curve then displays the part of the Schwarzschild branch that is unphysical for the Schwarzschild--BR black hole.   
}
\label{Fig1}
\end{figure}

\begin{figure}
    \centering
\includegraphics[width=0.45\textwidth]{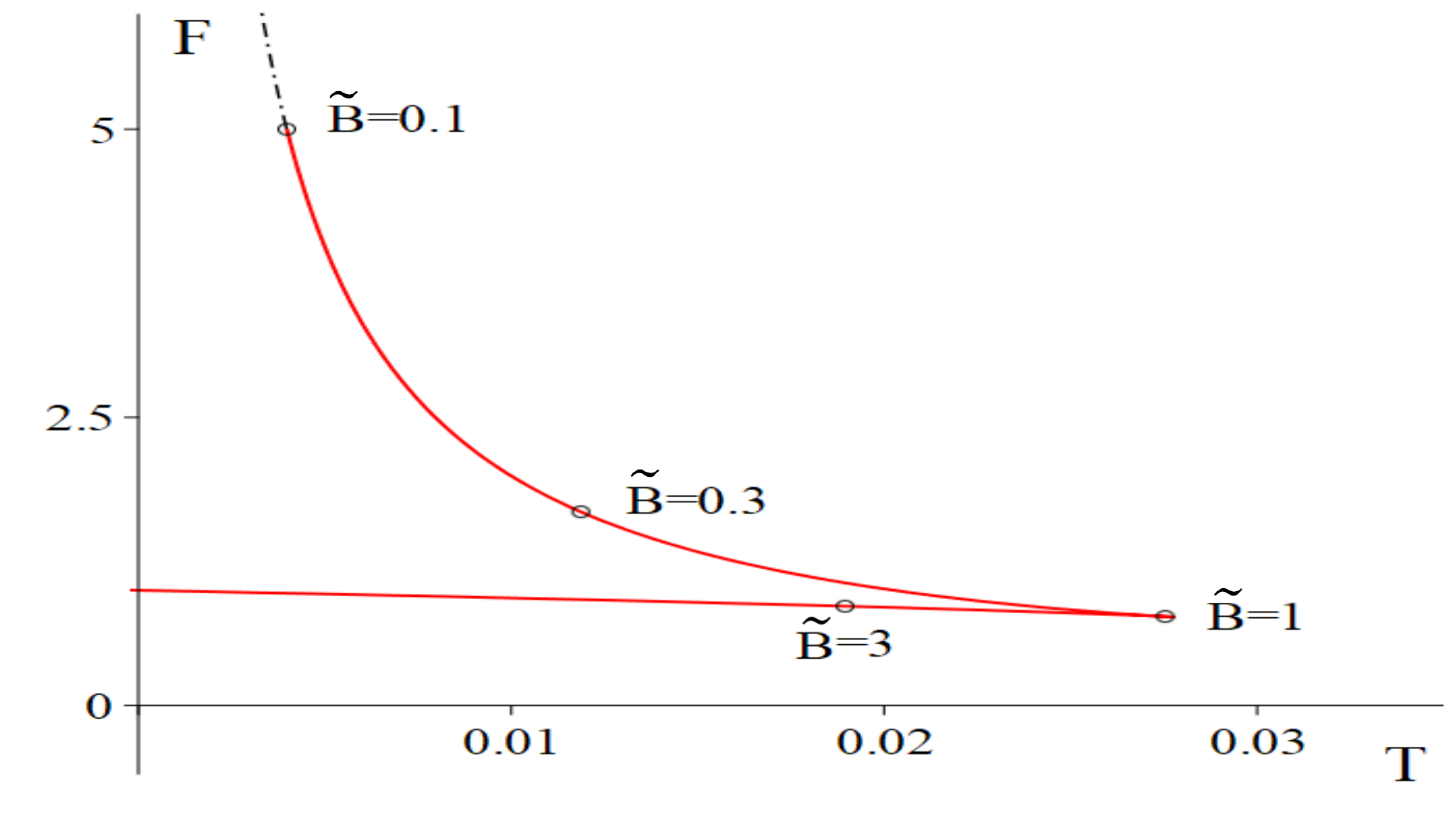}
\caption{{\bf Free energy $F$ of the Kerr--BR black hole}, displayed for several values of ${\tilde B=3, 1, 0.3, 0.1}$, and ${J=1}$. Similar to the Schwarzschild--BR case, large (slowly rotating) black holes are {\em not allowed} by the magnetic field, and the corresponding free energy curve (emanating on the lower left from the extremal black hole, with ${M = \tilde{a} = \tilde{r}_+ }$ and thus ${T=0}$) terminates at a corresponding  black circle. 
}
\label{Fig2}

\end{figure}

In the non-rotating ${J=0}$  case, the above formulas simplify drastically, and we obtain a simple expression
\be 
F=\frac{M}{2}=\frac{1}{16\pi\, T}\,.
\ee 
The constraint \eqref{constr} gives us a restriction on the admissible size of the hole, ${r_+\leq 2/\tilde{B}}$, and the implied restrictions on the free energy and temperature: ${F\leq 1/(2\tilde{B})}$ and ${T\geq \tilde{B}/(8\pi)}$.
Thus, the presence of a magnetic field forbids the existence of large black holes with small temperatures. 
The corresponding $F-T$ diagram is displayed in Fig.~\ref{Fig1}, where, for a given $\tilde B$, we observe Schwarzschild-like behavior, with a termination point (displayed by a black circle) determined by the constraint.

For the rotating case, the behavior of ${F=F(T,J)}$ must be plotted parametrically. This is displayed in Fig.~\ref{Fig2}  for ${J=1}$ and various values of $\tilde{B}$. We observe Kerr-like behavior with two branches that join at a cusp (occurring at the maximal admissible temperature). The lower branch of (small) near-extremal black holes is thermodynamically preferred.  
Similar to the Schwarzschild--BR case, for a given $\tilde{B}$, the constraint \eqref{constr} cuts off 
large (slowly rotating) black holes beyond a termination point displayed by black circles.

\subsection{Grandcanonical ensemble}

\begin{figure}
    \centering
\includegraphics[width=0.45\textwidth]{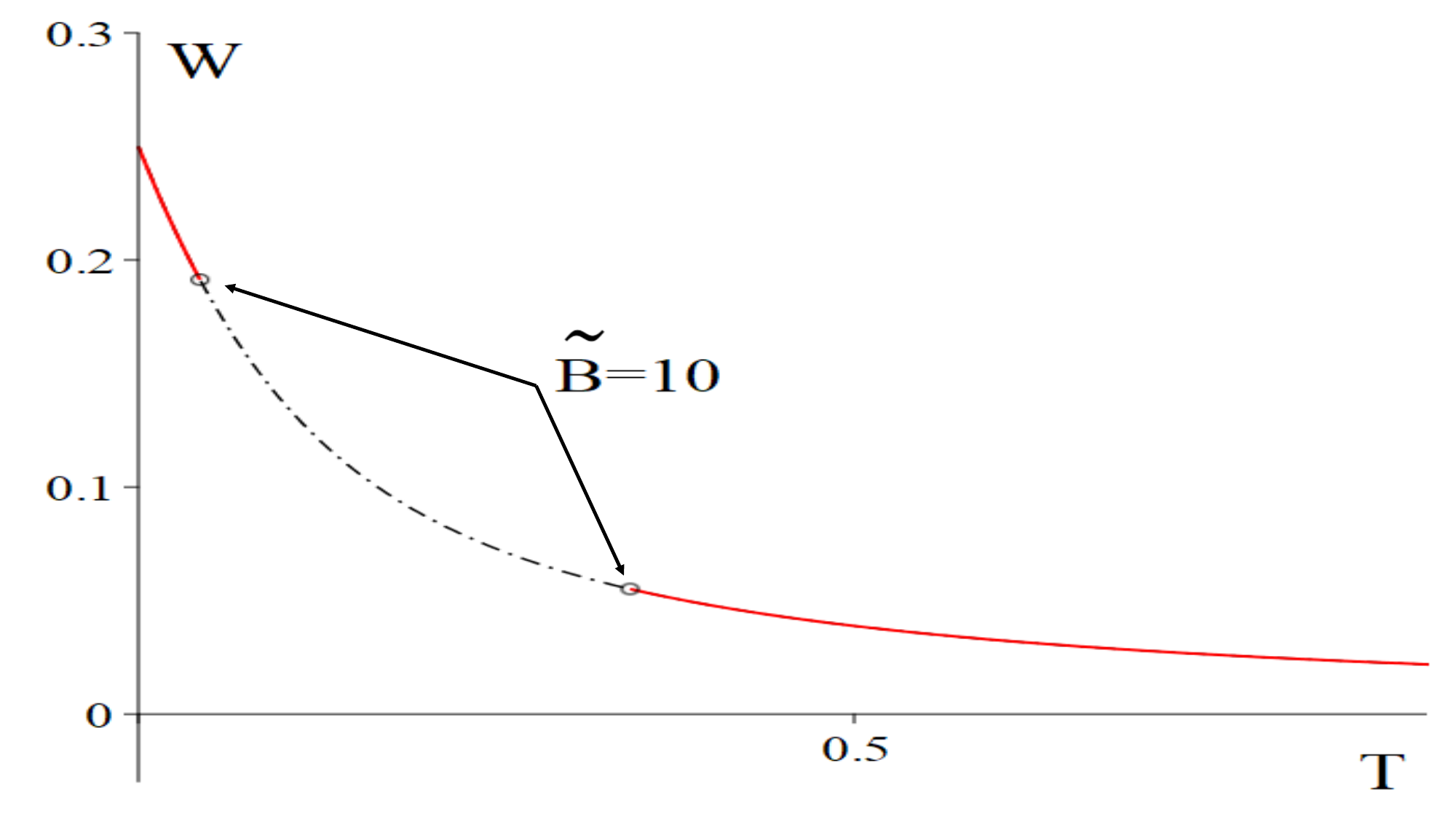}
\caption{{\bf Grandcanonical free energy $W$ of the Kerr--BR black hole} is displayed for $\tilde{B}=10>\tilde{B}_c$ and ${\Omega_+=1}$. Since ${\tilde{B}>\tilde{B}_c}$, the mid-size (average-rotating) black holes denoted by dashed black curve are excluded. The admissible large (fast spinning) black holes correspond to left red curve, whereas the small (slowly-spinning) ones are displayed by the right red curve, both terminating at the corresponding black circles. In the same diagram we also plot the case ${\tilde{B}<\tilde{B}_c}$, for which the entire curve is admissible. 
}
\label{Fig3}
\end{figure}

In the grandcanonical (fixed $\Omega_+$) ensemble, we have the following free energy 
\ba 
W \!\!& \equiv &\!\! W(T,\Omega_+)
=M-TS-\Omega_+ J\nonumber\\
\rovno \frac{M}{2}=\frac{\tilde{r}_+^2+\tilde{a}^2}{4\tilde{r}_+}\,,
\ea 
with $T$ and $\Omega_+$ given parametrically by \eqref{KerrTDs}. Assuming ${0\leq \tilde{a}\leq \tilde{r}_+}$ and inverting $\Omega_+$, we get 
\be
\tilde{a}=\frac{1}{2\Omega_+}\Big(1-\sqrt{1-4\,\tilde{r}_+^2\Omega_+^2}\,\Big)\,,
\ee
giving a constraint ${\tilde{r}_+\in [0, {\tilde{r}}_{\mbox{\tiny max}}\equiv 1/(2\Omega_+)]}$. Interestingly, in this range of $\tilde{r}_+$, the expression ${(M^2-\tilde{a}^2)}$ appearing in the constraint \eqref{constr} assumes its maximum at ${\tilde{r}_+=\tilde{r}_c\equiv\sqrt{2(\sqrt{5}-1)}/(4\Omega_+)\approx 0.393/\Omega_+}$. This results in a critical value of the magnetic field 
\be 
\tilde{B}_c\equiv\Omega_+\,\sqrt{10\sqrt{5}+22}\approx 6.660\, \Omega_+\,.
\ee 
For ${\tilde{B}\geq \tilde{B}_c}$,  the constraint \eqref{constr} is non-trivial and forbids mid-size average-rotating black holes, as displayed in the $W-T$ diagram in Fig.~\ref{Fig3} for  ${\tilde{B}=10>\tilde{B}_c}$ and ${\Omega_+=1}$. On the other hand, for ${\tilde{B}<\tilde{B}_c}$, the constraint \eqref{constr} is automatically satisfied, and all Kerr-like black holes in the grandcanonical ensemble thus exist. 

\section{Summary}

In this work we have proposed a consistent thermodynamic description of a novel genuine Kerr--BR black hole solution describing an uncharged rotating black hole immersed in external magnetic field \cite{OPnew}. Remarkably, upon a proper rescaling of the coordinates and parameters of the solution, the corresponding thermodynamic quantities   formally coincide with those of the vacuum Kerr spacetime. Consequently, they satisfy the standard CR formula, along with the standard first law and Smarr relation. However, the corresponding thermodynamic behavior is affected by the presence of a non-trivial parameter constraint \eqref{constr}, and  significantly differs from that of Kerr in both canonical  and grandcanonical ensembles.

Despite the appealing simplicity of the obtained thermodynamic description, there are several potential issues with its derivation that should be kept in mind. 
Namely, 

i) Given the highly non-trivial asymptotics, can one really use the upgraded Komar integration to calculate the thermodynamic mass and angular momentum? 

ii) Even if so, and taking into account the recent analysis of the global structure of the original Kerr--BR solution \cite{Zhou:2026tkm}, what is the most appropriate choice of the co-dimension two hypersurface over which such integrals should be evaluated?  

iii) Were we right to ignore the potential asymptotic contribution $\Omega_{\mbox{\tiny inf}}$ to the horizon velocity? 

iv) Most importantly, in our considerations, we have fixed the normalization factor $\beta$ of the timelike Killing vector by imposing the {\em standard} first law (or, alternatively, the standard CR formula). Consequently, the resultant first law has higher than full cohomogeneity. Should we instead, to maintain the standard full cohomogeneity,  demand the presence of the $\mu\, \delta B$ term \eqref{mudB} ${\grave{a}\,\, la}$ \cite{Gibbons:2013dna}, and consider the following extended first law and Smarr relations:
\begin{align}
    \delta \mathcal{M}&=T\,\delta S+(\Omega_+-\Omega_{\mbox{\tiny inf}}) \delta {J}-\mu\, \delta B \,,\label{first_law_modMT}\\
    \mathcal{M}&=2\,TS+2\,(\Omega_+-\Omega_{\mbox{\tiny inf}}){J}+\mu\, B\,,\label{Smarr_modMT}
\end{align}
resulting potentially in a different normalization factor $\beta$ and a modified CR formula?

v) Finally, it remains to be seen whether the present analysis can also be extended to the recently discovered
{\em charged} Kerr--Newman--Bertotti--Robinson black hole solution   \cite{OPnew}. If so, can the thermodynamics constructed in  \cite{Hu:2026slp} be recovered as its special case?

\subsection*{Acknowledgements}
D.K. acknowledges support from the
Charles University Research Center Grant No.
UNCE24/SCI/016.
J.P. and H.O. are grateful to the Czech Science Foundation Grant No. GA\v{C}R 26-22381S. H.O. also acknowledges the Charles University Grant No. GAUK 260325.

\appendix

\section{Upgraded Komar integration: mass and angular momentum}\label{App}

In this appendix, we describe the procedure for calculating Komar charges for the genuine Kerr--BR black hole. The issue with Komar charges for spacetimes with a non-trivial electromagnetic field is hidden in the fact that the usual Komar 2-form $*d\xi$ is no longer necessarily closed. However, one can modify the Komar 2-form in such a way that the closeness is restored.
Namely, following \cite{Liu2022}, the modified Komar 2-form is given by
\begin{align}
    {Q}(\xi)=\dfrac{1}{4}\epsilon_{\mu\nu\rho\sigma}\,Q^{\mu\nu}(\xi)\,\dd x^{\rho}\wedge \dd x^{\sigma}\,,
\end{align}
where $\epsilon_{\mu\nu\rho\sigma}$ is the Levi-Civita tensor, 
\begin{align}
    Q^{\mu\nu}(\xi)=-2\nabla^{[\mu}\xi^{\nu]}-2F^{\mu\nu} A_{\sigma}\xi^{\sigma}-2\tilde{F}^{\mu\nu} \tilde{A}_{\sigma}\xi^{\sigma}\,,
\end{align}
$\xi$ is a Killing vector, and $\tilde {A}$ is the 1-form potential for the dual Maxwell tensor, $\tilde{{F}}=\dd\tilde A$. We then have 
$\dd {Q}(\xi)=0$.

In particular, for the gauge chosen in (\ref{A_1_form}), the angular momentum and mass, calculated via
\begin{align}
    M=\dfrac{1}{8\pi}\int_{\Sigma_{2}^{\infty}}{Q}(\partial_t)\,,\quad
    J=-\dfrac{1}{16\pi}\int_{\Sigma_2^{\infty}}{Q}(\partial_\phi)\,,\label{MJ_def}
\end{align}
give
\begin{align}
    M=\dfrac{m}{K\,\beta}\,,\qquad
    J=\dfrac{m\,a}{K^2}.
\end{align}
Interestingly, these results remain invariant under performing  
a gauge transformation
\begin{align}
    {A}\to {A}+{A}^0\,,\qquad \tilde{A}\to \tilde{A}+\tilde{A}^0\,,
\end{align}
with arbitrary {\em constant} 1-forms ${A}^0=A_t^0\,\dd t+A_{\phi}^0\, \dd \phi$ and $\tilde{A}^0=\tilde{A}_t^0\,\dd t+\tilde{A}_{\phi}^0\, \dd \phi$.

Moreover, integrating $\dd {Q}(\xi)=0$ and using Gauss' law, we obtain 
\begin{align}
    \dfrac{1}{8\pi}\int_{\Sigma_2^+}{Q}(\xi)=\dfrac{1}{8\pi}\int_{\Sigma_2^{\infty}}{Q}(\xi)\,.
\end{align}
Identifying $\xi$ with the Killing vector defined in (\ref{xi_def}), and using the definitions of mass and angular momentum \eqref{MJ_def}, the latter equation becomes:
\begin{align}
    M=\dfrac{1}{8\pi}\int_{\Sigma_2^+}{Q}(\xi)+2\,\Omega_+ J\,.
\end{align}
Calculating the corresponding integral {\em on the horizon} gives $2 TS$. We have thus directly recovered the  Smarr formula \eqref{Smarr}.


%

\end{document}